\documentclass[a4paper]{article}
\usepackage{graphicx}
\usepackage{amsmath}

\begin{document}

\begin{center}

{\bf \large \centering Reshuffling spins with short range interactions:\\
When sociophysics produces physical results}

\bigskip

A.O. Sousa$^{a,}$\footnote{\tt sousa@ica1.uni-stuttgart.de},
K. Malarz$^{b,}$\footnote{\tt malarz@agh.edu.pl}
 and  
S. Galam$^{c,}$\footnote{\tt galam@shs.polytechnique.fr}

\bigskip

{\em
$^a$ Institute for Computer Physics (ICP),
University of Stuttgart,\\
Pfaffenwaldring 27, D-70569 Stuttgart, Euroland\\[2mm]

$^b$ Faculty of Physics and Applied Computer Science,\\
AGH University of Science and Technology,\\
al. Mickiewicza 30, PL-30059 Krak\'ow, Euroland\\[2mm]

$^c$ Centre de Recherche en \'Epist\'emologie Appliqu\'ee,\\
CREA - \'Ecole Polytechnique, CNRS UMR 7656,\\
1, rue Descartes, F-75005 Paris, Euroland\\[2mm]
}

\today
\end{center}


\begin{abstract}
Galam reshuffling  introduced in opinion dynamics models is investigated under the nearest neighbor Ising model on a square lattice using Monte Carlo simulations.
While the corresponding Galam analytical critical temperature $T_C\approx 3.09$ $[J/k_B]$ is recovered almost exactly, it is proved to be different from both values, not reshuffled ($T_C=2/\text{arcsinh}(1)\approx 2.27$ $[J/k_B]$) and mean-field ($T_C=4$ $[J/k_B]$). 
On this basis, gradual reshuffling is studied as function of $0\leq p \leq 1$ where $p$ measures the probability of spin reshuffling after each Monte Carlo step.
The variation of $T_C$ as function of $p$ is obtained and exhibits a non-linear behavior.
The simplest Solomon network realization is noted to reproduce Galam $p=1$ result.
Similarly to the critical temperature, critical exponents are found to differ from both, the classical Ising case and the mean field values.
\end{abstract}



\section{Introduction}

Sociophysics is based on the use of concepts and tools from physics to describe social and political behavior \cite{strike}. While the validity of such a transfer has been long questioned among physicists \cite{testimony}, none ever has expected that some basic sociophysics question may in turn lead to new development within physics. In this paper we report, to our knowledge for the first time, on such a development. Indeed economics long ago has contributed substantially to statistical physics, but indirectly, providing several tools it developed on its own, which were then taken over by physics \cite{econo}.

Opinion dynamics has been a very active field of research among physicists in the recent years \cite{opiniondyn,weron,sznajd}.
Most models are based on some local update rule and corresponding results are obtained using numerical simulations \cite{opiniondyn,weron,sznajd}.
The Sznajd model initiated in 2000 \cite{weron} is the most popular and has generated a great deal of papers \cite{sznajd}.
In parallel, a much older one proposed by Galam in 1986 \cite{voting-old}, has stayed confined to only few works \cite{mino}. But a recent universal scheme for local rules shows that both models have the same dynamics \cite{unify} prompting the question of why they have been perceived as different with an advantage for the first one.

Indeed, the reason lies in the fact that, although the Galam model is solvable analytically, its very formulation has been perceived as the signature of an intrinsic mean field nature. Since its dynamics  is monitored by a total reshuffling of agents between repeated local updates, in principle everybody can interact to everybody. This fact has been understood as  everybody does interact to everybody simultaneously, as in a mean field treatment \cite{espagnol}. However that is not the case due to the local range of interactions which are restricted to separate small groups of agents after each reshuffling. At least, for years,  Galam was adamant in claiming it.

In this work we address this question  by a thorough numerical Monte Carlo  investigation  of the effect of reshuffling spins on the phase diagram of  the two-dimensional nearest neighbor (NN) Ising model \cite{ising}.
Reshuffling is introduced gradually according to the variable  $0\leq p \leq 1$ where $p$ is the probability of reshuffling all the spins of the lattice at each Monte Carlo step.
We call it the Gradually Reshuffled Ising Model and denote it by GRIM.
It is worth to stress that after each spin reshuffling, interactions stay local among NN.

The critical temperature $T_C$ is calculated for a series of  values of $p$ from $p=0$ (square lattice Ising model --- SLIM) up to $p=1$ (totally reshuffled Ising model --- TRIM).
Binder's cumulant for $T_C$ evaluation is used to avoid finite size effect \cite{binder}.
As expected, at $p=0$ the SLIM exact value $T_C=2/\mathrm{arcsinh}(1)\approx 2.27\, [J/k_B]$ is recovered. The variation of the GRIM $T_C$ as function of $p$ is found to exhibit a non-linear behavior. A similar study of gradual reshuffling was performed for Galam opinion model in the case of local groups of size four \cite{chopard1}.

The unifying Galam scheme \cite{unify} is then used to map the TRIM into a local rule model where update groups are of a size five. Its allows an analytical  calculation of the  critical temperature $T_C$. In  contrast to the simulations, it is found to depend on the choice of the dynamics.
Metropolis and  Glauber yield different values, respectively $T_C\approx 1.59\, [J/k_B]$ and $T_C\approx 3.09\, [J/k_B]$.
Last one reproduces almost exactly the Monte Carlo result at $p=1$, $T_C\approx 3.03\, [J/k_B]$.
The simplest realization \cite{solomon} of the Solomon network (SN) \cite{erez} is noted to reproduce TRIM results.
Similarly to the critical temperature, critical exponents are found to differ from both, the SLIM case and the mean field values. Concluding remarks with respect to future work end the paper.

\section{The gradually reshuffled Ising model}

We consider the nearest neighbor Ising model \cite{ising} on a square lattice
with ferromagnetic interactions
\begin{equation}
\label{eq-ham}
{\cal H}= - \dfrac{1}{2} \sum_{i,j} J_{ij} S_i S_j,
\end{equation}
where $S_i = \pm 1$ is the Ising spin variable at each node $i$ of the square lattice and
\[
J_{ij}=\begin{cases}
J>0 & \text{if $i$ and $j$ are neighbors,} \\
0   & \text{otherwise.}
\end{cases}
\]
is short-range ferromagnetic exchange integral.

Monte Carlo simulations are performed using either a Glauber or a Metropolis
dynamics. In the Glauber dynamics, every time step all spins at the lattice are
investigated in type-writer fashion, i.e. spin-by-spin: from left to right and
then from top to bottom. For each spin $i$ in an initial configuration $\mu_i$,
a new configuration $\eta_i$ resulting from the single spin flip $S_i\to -S_i$
is accepted with a probability
\begin{equation}
\label{proba-g}
p^G_{\mu_i \to \eta_i}=\dfrac{\exp(-E_{\eta_i}/k_BT)}{\exp(-E_{\mu_i}/k_BT)+\exp(-E_{\eta_i}/k_BT)},
\end{equation}
where $E_{\eta_i}$  is the energy of configuration $\eta_i$, $E_{\mu_i}=-E_{\eta_i}$  is the energy of configuration $\mu_i$ and $k_B$ is the Boltzmann constant. When all $N$ spins are
investigated one Monte Carlo step (MCS) is completed. In Metropolis scheme,
the acceptance probability of the new configuration may be expressed in a
simple form
\begin{equation}
\label{proba-m}
p^M_{\mu_i \to \eta_i}=\min\{ 1, \exp [ -(E_{\eta_i}-E_{\mu_i})/k_BT ] \}.
\end{equation}
But here, at  each MCS, all the spins are randomly visited and updated (a random list of
spins labels assures that each spin is reached exactly once). 

The value of the critical temperature $T_C$ is extracted from the evaluation of the order parameter  $m=\sum_i S_i/N$ as a function of $T$.  In a second step, to get a more precise estimate of $T_C$ in the thermodynamic limit, we calculate Binder's fourth cumulant of the order parameter defined by
\begin{equation}
U\equiv 1-\frac{ \langle m^4 \rangle }{ 3 \langle m^2 \rangle ^2 },
\label{eq-bc}
\end{equation}
where $\langle\cdots\rangle$ represent the thermal average (taken over 400000
MCS after discarding 100000 MCS for equilibration at each temperature). This
cumulant should go to $2/3$ below $T_C$ and to zero above $T_C$ when the size
increases, and the finite-size estimates are expected to cross around $T_C$ 
\cite{binder}.

Our Monte Carlo simulations have been performed on square lattices of size 
$50\le L\le 500$ assuming semi-periodic (helical) or periodic boundary conditions.
For given temperature $T$, simulation starts with $m=1$, i.e., with all spins $S_i=1$.
In every MCS, before updating all spins, with a probability $p$ the
reshuffling procedure takes place: random permutation of all spin labels is
produced and their positions are rearranged according to that new labels
order.
With probability $1-p$, all spins keep their current position.

\section{The results}

\begin{figure}
\includegraphics[width=.9\textwidth]{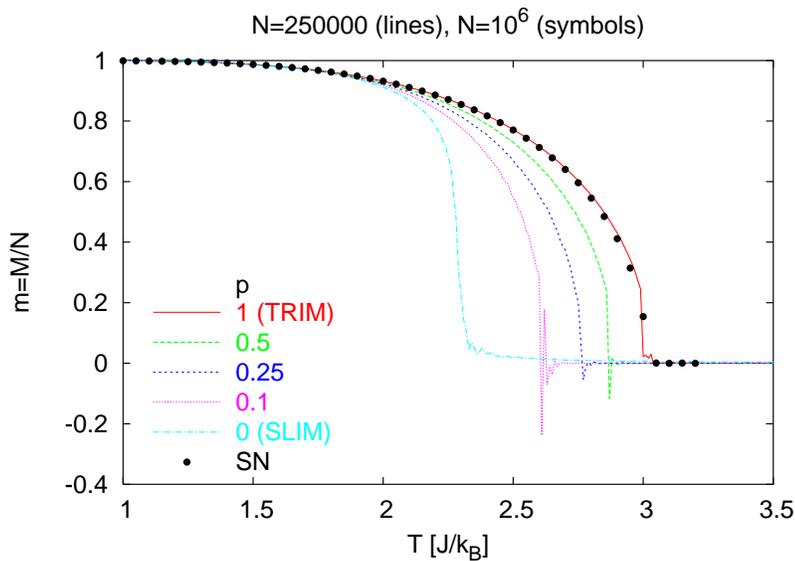}
\caption{Magnetization $m(T)$ dependence for $L=500$ obtained with Glauber dynamics. 
For the SN $N=10^6$ spins were simulated.}
\label{fig-1}
\end{figure}

In Fig. \ref{fig-1} $m(T)$ for size $L=500$ and different reshuffling probabilities $p$ are shown.
Also $m(T)$ for SN browsed from Ref. \cite{solomon} for a million spins is included.
The latter agrees very well with reshuffling case for $p=1$ (TRIM).
In SN a random permutation of spin labels is created and each spin interacts with its two NN and two additional neighbors of its mirror in reshuffled lattice.
In contrast to current work the reshuffling procedure takes place only once: before the simulation starts and the spin neighborhood is fixed during simulation.
The spin dynamics is governed via Eqs. \eqref{eq-ham}, \eqref{proba-g} and \eqref{proba-m} but links are directed.
It means, that if spin $j$ is a neighbor of spin $i$ does not yield $i$ being neighbor of $j$.

\begin{figure}
\includegraphics[width=.9\textwidth]{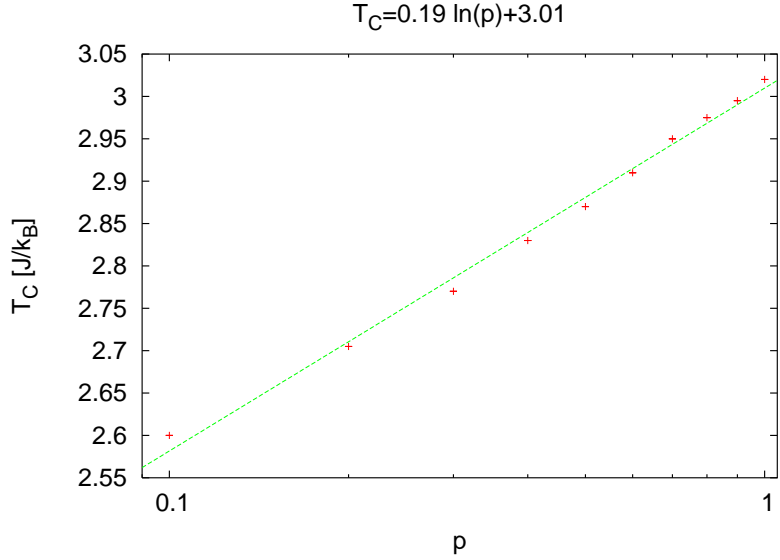}
\caption{Dependence of Curie temperature $T_C(p)=0.19\ln p+3.01$ on the reshuffling probability $p$ given via cross-point of $U(T)$ curves for different system sizes $50\le L\le 300$.}
\label{fig-3}
\end{figure}

Decreasing reshuffling probability $p$ shifts $m(T)$ towards SLIM with $T_C=2/\mathrm{arcsinh}(1)\approx 2.27\, [J/k_B]$ \cite{onsager}.
The dependence $T_C(p)$ is found to be nonlinear as a function of the reshuffling parameter $p$.
In the range $p>0.1$ it may be approximated with the logarithmic law $T_C=0.19\ln p+3.01$ (see Fig. \ref{fig-3}).

\begin{figure}
\includegraphics[width=.9\textwidth]{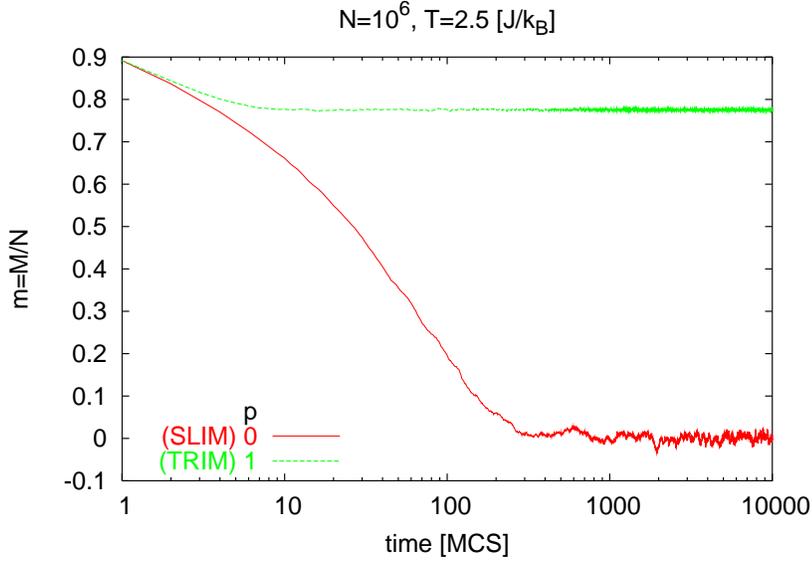}
\caption{Time evolution of the magnetization $m(T)$ obtained with Glauber dynamics at $T=2.5\, [J/k_B]$ for $p=0$ and $p=1$.
Associated $T_C$ are respectively $T_C=2.27\, [J/k_B]$ and $T_C=3.03\, [J/k_B]$.
Simulations are carried out for $N=10^6$ spins.
}
\label{fig-2}
\end{figure}

Examples of the order parameter time evolution $m(T)$ at fixed temperature $T=2.5\, [J/k_B]$ are shown in Fig. \ref{fig-2} for $p=0$ (SLIM) and $p=1$ (TRIM).
The curves show that $m(T)$ vanishes for $p=0$ as expected from its $T_C=2.27<2.5\, [J/k_B]$.
On the opposite for $p=1$ we have $T_C=3.03>2.5\, [J/k_B]$ and $m(T)$ stabilizes at a non zero value as it should be in the associated ordered phase.


\begin{figure}
\includegraphics[width=.9\textwidth]{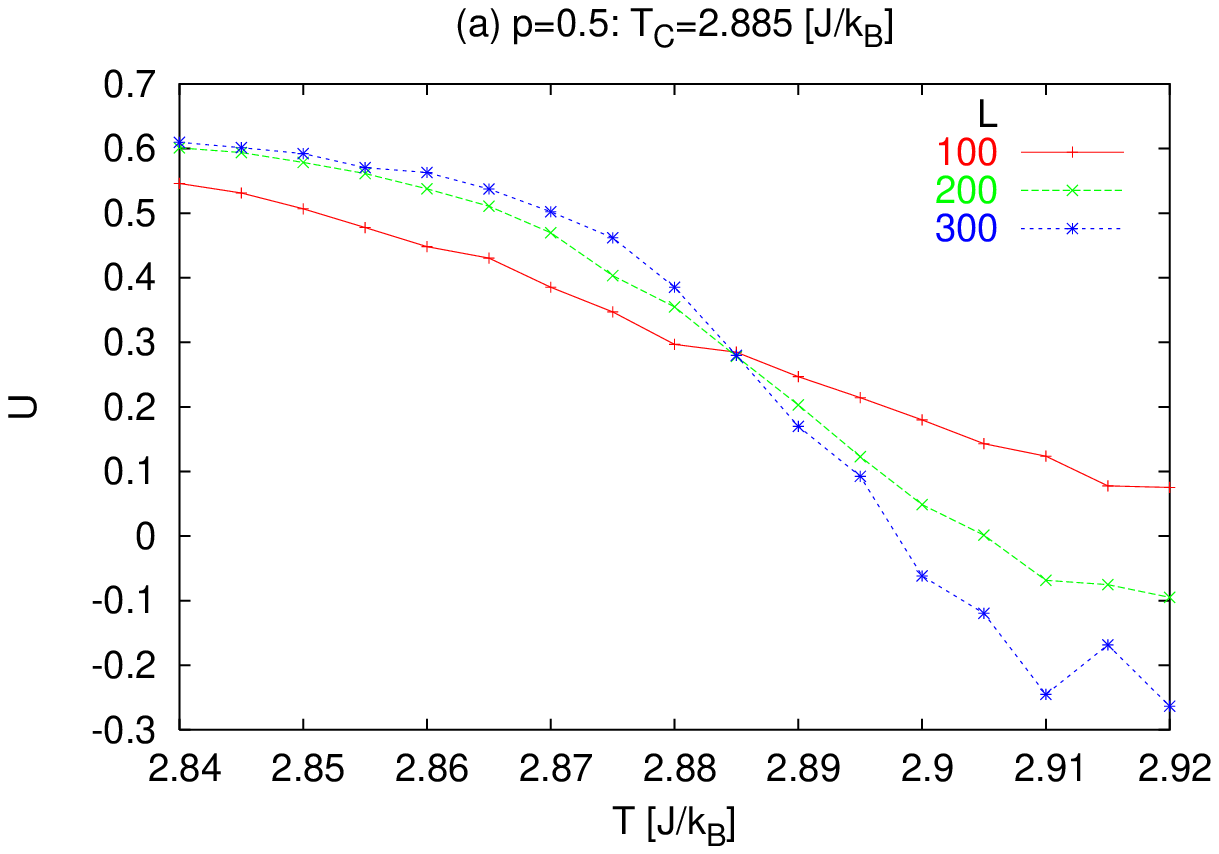}\\
\includegraphics[width=.9\textwidth]{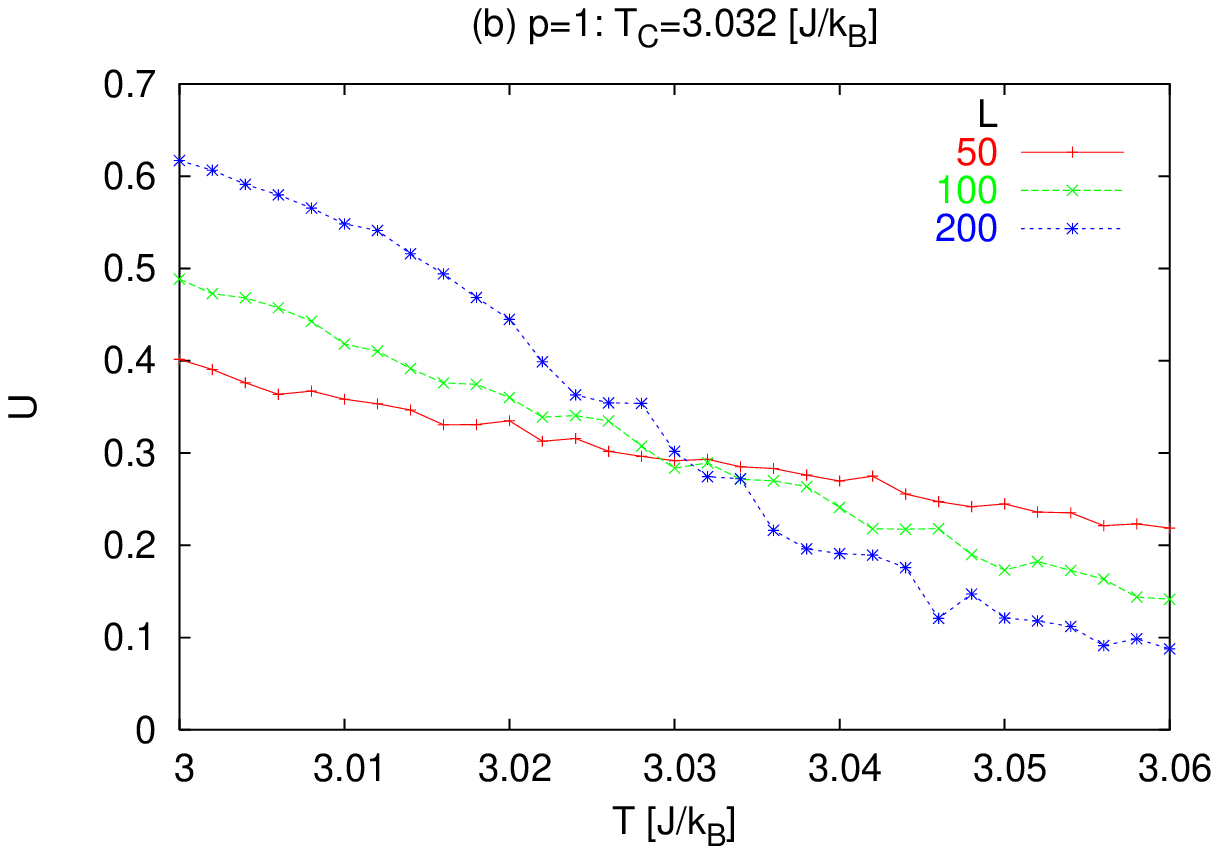}
\caption{Binder's cumulant $U$ dependence on the temperature $T$ for different system sizes $L$ and reshuffling probability (a) $p=1/2$ and (b) $p=1$.}
\label{fig-4}
\end{figure}

In Fig. \ref{fig-4} Binder's cumulant $U$ dependence on the temperature $T$ for different system sizes $L$ and reshuffling probability $p$ is presented.
The common crossing-point of these curves predicts Curie temperature $T_C$.

\begin{figure}
\includegraphics[width=.9\textwidth]{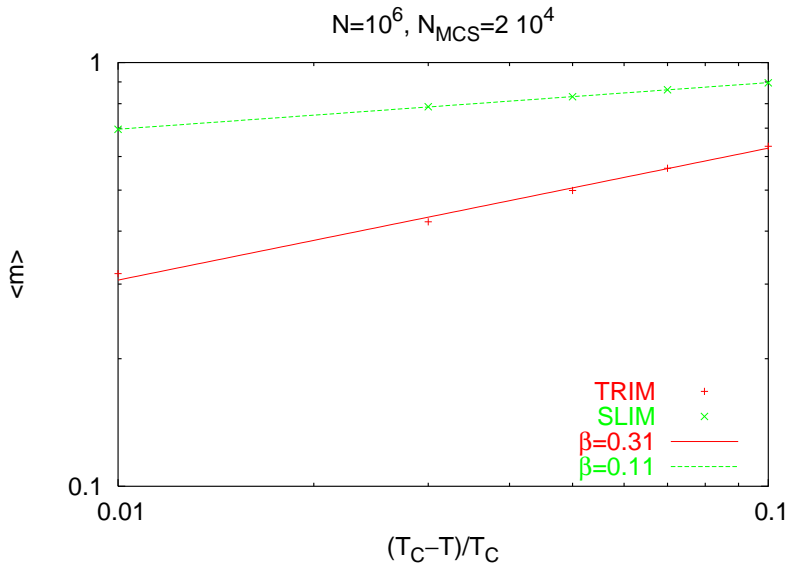}
\caption{Critical exponents $\beta$ --- which describes magnetization behavior in the vicinity of $T_C$ --- for $p=0$ and $p=1$ are $\beta_{\text{SLIM}}=0.11$ and $\beta_{\text{TRIM}}=0.31$, respectively.
$N=10^6$ spins were simulated.
$N_{\text{MCS}}=2\cdot 10^4$ and $\langle m\rangle$ is averaged over last $10^4$ [MCS].
}
\label{fig-5}
\end{figure}

In Fig. \ref{fig-5} critical exponents $\beta$ --- which describe magnetization behavior in the vicinity of $T_C$ --- are presented.
For $p=0$ and $p=1$ these exponents are respectively $\beta_{\text{SLIM}}=0.11$ and $\beta_{\text{TRIM}}=0.31$.
The exact value at $p=0$ is $\beta^{\text{th}}_{\text{SLIM}}=1/8$ \cite{onsager,krytyczne} while the mean field value is $\beta^{\text{th}}_{\text{MF}}=1/2$ \cite{krytyczne}. 
From $\beta_{\text{TRIM}}=0.31 \neq 1/2=\beta^{\text{th}}_{\text{MF}}$, we can conclude that TRIM, and thus  Galam reshuffled models,  are not mean field models, since  otherwise the associated critical exponents  would have been mean field exponents, as it must be for any mean field approximation.

\section{The Galam unifying scheme}

We now go one step further in the investigation of the validity of
Galam unifying scheme  \cite{unify}.  It is a general sequential
frame, which  operates via local updates of small groups of spins
randomly selected. Given one group, a specific probabilistic majority
rule is then applied for  each peculiar  configuration. It  is a
function of the local value of the order parameter, i.e., the ratio of
majority to minority.

Applying this scheme to the square NN Ising model investigated above
and following the Monte Carlo procedure, which considers one spin and
its four NN leads to use local groups of one  size, with five spins. On
this basis, we define $g_k$ as  the probability that a group of five
spins with $k$ plus and $(5-k)$ minus ends up with five plus.
Simultaneously the probability to get five minus is $(1-g_k)$. With
five spins, the number $k$ of plus spins can vary from $5$ down to $0$
producing six different coefficients $g_k$. But from up-down symmetry this
number reduces to three with  $g_{0}=1-g_{5}$, $g_{1}=1-g_{4}$ and
$g_{2}=1-g_{3}$. On this basis,  if $p(t)$ is the proportion of plus
spins at time $t$, we obtain for the new proportion $p(t+1)$  after one
update cycle
\begin{equation}
\label{eq-ptp1}
p(t+1)=\sum_{k=0}^5 {5 \choose k} g_k [p(t)]^k [1-p(t)]^{5-k},
\end{equation}
which is a binomial with ${5 \choose k}=\frac{5!}{5!(5-k)!}$.

To evaluate the coefficients $g_{i}$ in Eq. \eqref{eq-ptp1} we enumerate all possible $2^5=32$ configurations associated to a local group of size five and regroup them by their respective number of plus spins.
The corresponding energies $E_{\mu_i}$ are calculated as well as energies $E_{\eta_i}$ are obtained once the central spin has been flipped.
As described in Sect. 2 the central spin flip is accepted with probability given by either Eq. \eqref{proba-g} or \eqref{proba-m} depending on chosen dynamics, Glauber or Metropolis. 
However, here  this probability is weighted by the plus spin proportion \cite{unify}.
We now illustrate the scheme in one case with four plus spins ($S_i=+1$) and one minus ($S_i=-1$).
Being on a square lattice, either the minus spin is at the center or not.
One configuration corresponds to the first case,
\[
\left(\begin{array}{ccc} & + &  \\+ & - & + \\ & + & \end{array}\right)\rightarrow
\left(\begin{array}{ccc} & + &  \\+ & + & + \\ & +& \end{array}\right), 
\]
and for the second case,  there exist four configurations  of the type
\[
\left(\begin{array}{ccc} & + &  \\+ & + & + \\ & - & \end{array}\right)\rightarrow
\left(\begin{array}{ccc} & + &  \\+ & - & + \\ & - & \end{array}\right).
\]
where the minus spin can rotate on each one of the four NN sites. 

Using Glauber dynamics, in the first case with one configuration, the flip is validated with a probability $a$ weighted by with a proportion one of plus spins.
In contrast, the initial configuration is conserved with a probability $(1-a)$ weighted by $4/5$,  the proportion of plus spin.
In the second case, the probability for a flip is given  by $b$ weighted with a proportion $3/5$ of plus spins while for no flip we have  respectively $(1-b)$  and $4/5$, both values being multiplied by four.
>From Eq. \eqref{proba-g} we have
\begin{equation}
a=\dfrac{ \exp(4J/k_BT) }{ \exp(4J/k_BT)+\exp(-4J/k_BT) },
b=\dfrac{ \exp(-2J/k_BT) }{ \exp(2J/k_BT)+\exp(-2J/k_BT) }.
\end{equation}
Combining above results yields for  the coefficient $g_5$ in Eq. (5),
\begin{equation}
g_5= \frac{4+a}{5}.
\end{equation}
In a similar fashion, performing similar above evaluation for the other configurations  we get the coefficients $\{g_i\}$ with
\begin{equation}
g_4= \frac{20+a-4b}{25}, 
\qquad
g_3= \frac{31-4b}{50}.
\end{equation}

In parallel, using Metropolis dynamics from Eq. \eqref{proba-m} the 
first configuration leads to a flip with probability one ($\min[1,\exp(8J/k_BT)]=1$), and the second one is performed with probability
\begin{equation}
c=\exp{(-4J/k_BT)}.
\end{equation}
The corresponding coefficients $\{g_i\}$ become
\begin{equation}
g_5= \frac{5-d}{5},
\qquad
g_4= \frac{21-4c}{25}
\qquad
\text{and}
\qquad
g_3= \frac{28}{50},
\end{equation}
where $d=\exp(-8J/k_BT)$
is a probability of the configuration change
\[
\left(\begin{array}{ccc} & + &  \\+ & + & + \\ & + & \end{array}\right)\rightarrow
\left(\begin{array}{ccc} & + &  \\+ & - & + \\ & +& \end{array}\right)
\]
given by Metropolis dynamics.

At this stage, we can calculate the unstable fixed points from  Eq. \eqref{eq-ptp1} to get the corresponding critical temperature, respectively  $T_C=3.09\, [J/k_B]$ for Glauber, and $T_C=1.59\, [J/k_B]$ for Metropolis. The Glauber result is rather close to the numerical finding  $T_C=3.03$ $[J/k_B]$ but yet a bit different. At present we do not understand why. In  contrast, the simulation results are invariant in the choice of the dynamics. It is worth to note that a value of $T_C$ a little bit larger than 3 $[J/k_B]$ was obtained previously by Malarz from a Monte Carlo simulation of SN \cite{solomon}. 

\section{ Discussion}

In this work, we have studied the effect of gradual spin reshuffling on the phase diagram of the lattice square Ising model. Performing a series of Monte Carlo simulations, we have proved that Galam reshuffling is neither a mean-field treatment nor the usual Ising model (SLIM). In particular it yields non mean field exponents. 

On this basis, new physical questions arise. First, does reshuffling create a new universality class for the Ising model? At which value of the reshuffling parameter $p$ does the crossover occur? Besides, the theoretical interest, and the sociophysics consequences, to find a physical situation which could correspond to reshuffling would be also interesting. 

\bigskip
\noindent
{\bf Acknowledgments.}
The authors are grateful to Dietrich Stauffer for many stimulating discussions and permanent criticism of their ideas, claims and results.
Part of numerical calculations was carried out in ACK\---CY\-F\-RO\-NET\---AGH.
The machine time on HP Integrity Superdome is financed by the Polish Ministry of Science and Information Technology under grant No. MNiI/\-HP\_I\_SD/\-AGH/047/\-2004.



\begin{thebibliography}{00}

\bibitem{strike}
D. Stauffer,
Comput. Sci. Eng. {\bf 5}, 71 (2003);

D. Stauffer,
Fractals {\bf 11}, 313 (2003);

D. Stauffer,
Physica {\bf A336}, 1 (2004);

S. Galam, Y. Gefen and Y. Shapir, 
Math. J. Socio. {\bf 9}, 1 (1982).  
 
\bibitem{testimony} S. Galam, 
Physica {\bf A336}, 49 (2004).
 
\bibitem{econo} V. Pareto,
{\em Cours d'Economique Politique}, Vol. 2,
(Macmillan, Paris, 1897).

\bibitem{opiniondyn}
D. Stauffer,
{\tt physics/0503115};
                                                                                
P. Gawro\'nski, P. Gronek, K. Ku{\l}akowski,
{\tt physics/0503085};
                                                                                
P. Gawro\'nski, P. Gronek, K. Ku{\l}akowski,
{\tt  physics/0501160};

S. Fortunato, 
{\tt cond-mat/0501105};
     
K. Ku{\l}akowski, P. Gawro\'nski, P. Gronek,
{\tt physics/0501073};

A.O. Sousa,
Physica {\bf A348}, 701 (2005);

S. Fortunato,
Physica {\bf A348}, 683 (2005);

S. Fortunato,
{\tt cond-mat/0405083};

E. Ben-Naim,
Europhys. Lett. {\bf 69}, 671 (2005);

M. F. Laguna, S. Risau-Gusman, G. Abramson, S. Goncalves, J. R. Iglesias,
{\tt nlin.AO/0404024};

M.F. He, Q. Sun, H.S. Wang,
Int. J. Mod. Phys. {\bf C15}, 767 (2004);

D. Stauffer, A.O. Sousa, C. Schulze,
{\tt cond-mat/0310243};

C.J. Tessone, R. Toral, P. Amengual, H.S. Wio, M. San Miguel,
Eur. Phys. J. {\bf B39}, 535 (2004);

D. Stauffer, H. Meyer-Ortmanns,
Int. J. Mod. Phys. {\bf C15}, 241 (2004); 

R. Hegselmann, U. Krause,
J. Artificial Soc. Soc. Simul.\footnote{available online at {\tt http://jasss.soc.surrey.ac.uk/}} {\bf 5} (3) 2 (2002);

G. Deffuant, D. Neau, F. Amblard, G. Weisbuch,
Complex Syst. {\bf 3}, 87 (2000);

G. Weisbuch, G. Deffuant, F. Amblard, J.-P. Nadal,
Complexity {\bf 7}, 55 (2002);

G. Deffuant, F. Amblard, G. Weisbuch, T. Faure,
J. Artificial Soc. Soc. Simul. {\bf 5} (4) 1 (2002);

J.A. Ho{\l}yst, K. Kacperski, F. Schweitzer,
Ann. Rev. Comput. Phys. {\bf IX}, 253 (2001) and references therein.

\bibitem{weron} K. Sznajd-Weron, J. Sznajd, 
Int. J. Mod. Phys. \textbf{C11}, 1157 (2000).

\bibitem{sznajd}
C. Schulze,
Int. J. Mod. Phys. {\bf C15}, 867 (2004);

M.F. He, B. Li, L.D. Luo,
Int. J. Mod. Phys. {\bf C15}, 997 (2004);

D. Stauffer,
J. Math. Sociol. {\bf 28}, 25 (2004);

L. Sabatelli, P. Richmond,
Physica {\bf A334}, 274 (2004);

J.J. Schneider,
Int. J. Mod. Phys. {\bf C15}, 659 (2004);

C. Schulze,
Int. J. Mod. Phys. {\bf C15}, 569 (2004);

L. Behera, F. Schweitzer,
Int. J. Mod. Phys. {\bf C14}, 1331 (2003);

L. Sabatelli, P. Richmond,
Int. J. Mod. Phys. {\bf C14}, 1223 (2003);

J. Bonnekoh,
Int. J. Mod. Phys. {\bf C14}, 1231 (2003);

F. Slanina, H. Lavicka,
Eur. Phys. J. {\bf B35}, 279 (2003);

C. Schulze,
Physica {\bf A324}, 717 (2003);

C. Schulze,
Int. J. Mod. Phys. {\bf C14}, 95 (2003);

D. Stauffer, P.M.C. de Oliveira,
Eur. Phys. J. {\bf B30}, 587 (2002);

D. Stauffer,
Comput. Phys. Commun. {\bf 146}, 93 (2002);

D. Stauffer,
Int. J. Mod. Phys. {\bf C13}, 315 (2002);

A.T. Bernardes, D. Stauffer, J. Kertesz,
Eur. Phys. J. {\bf B25}, 123 (2002);

A.S. Elgazzar,
Int. J. Mod. Phys. {\bf C12}, 1537 (2001);

I. Chang,
Int. J. Mod. Phys. {\bf C12}, 1509 (2001); 

R. Ochrombel,
Int. J. Mod. Phys. {\bf C12}, 1091 (2001);

A.T. Bernardes, U.M.S. Costa, A.D. Araujo, D. Stauffer,
Int. J. Mod. Phys. {\bf C12}, 159 (2001);

A.A. Moreira, J.S. Andrade, D. Stauffer,
Int. J. Mod. Phys. {\bf C12}, 39 (2001);

D. Stauffer, A.O. Sousa, S.M. de Oliveira,
Int. J. Mod. Phys. {\bf C11}, 1239 (2000).

\bibitem{voting-old}
S. Galam, 
J. Math. Psychology \textbf{30}, 426 (1986);

S. Galam, 
J. Stat. Phys. \textbf{61}, 943 (1990);

S. Galam, 
Physica  \textbf{A285}, 66 (2000).
 
\bibitem{mino} S. Galam,  
Eur. Phys. J. \textbf{B25}, 403 (2002).
 
\bibitem{unify} S. Galam,
{\tt cond-mat/0409484}.

\bibitem{espagnol}
C.J. Tessone, R. Toral, P. Amengual, H.S. Wio, M. San Miguel,
Eur. Phys. J. {\bf B39}, 535 (2004).

\bibitem{ising}
W. Lenz,
Z. Phys. {\bf 21} (1920) 613;

E. Ising,
Z. Phys. {\bf 31} (1925) 253.

\bibitem{binder} K. Binder, D.W. Heermann,
{\em Monte Carlo Simulation in Statistical Phyics},
(Springer Verlag, 1988).

\bibitem{chopard1} S. Galam, B. Chopard, A. Masselot, M. Droz,
Eur. Phys. J. \textbf{B4}, 529 (1998).
 
\bibitem{solomon} K. Malarz,
Int. J. Mod. Phys. \textbf{C14}, 561 (2003).

\bibitem{erez} T. Erez, M. Hohnisch, S. Solomon,
{\em Economics: Complex Windows},
Eds. M. Salzano and A. Kirman,
(Springer, 2005), p. 201,
[{\tt cond-mat/0406369}].

\bibitem{herrmann} M.C. Gonzalez, A.O. Sousa, H.J. Herrmann,
Int. J. Mod. Phys. \textbf{C15}, 45 (2004).

\bibitem{stauffer-meyer} D. Stauffer, H. Meyer-Ortmanns,
Int. J. Mod. Phys. \textbf{C15}, 241 (2004).

\bibitem{slanina} F. Slanina, H. Lavicka,
Eur. Phys. J. \textbf{B35}, 279 (2003).

\bibitem{frank} F. Schweitzer, J. Zimmermann, H. M\"uhlenbein, 
Physica \textbf{A303}, 189 (2002).

\bibitem{onsager} L. Onsager,
Phys. Rev. {\bf 65}, 117 (1944).

\bibitem{krytyczne} J.J. Binney, N.J. Dowrick, A.J. Fisher, M.E.J. Newman,
{\em A theory of critical phenomena.
An Introduction to the renormalization group.}
(Clarendon Press, Oxford, 1992) 

\end{thebibliography}
\end{document}